\documentclass{jpsj2}

\title{
Importance of Initial Electronic State in Two-Dimensional Electron System for Electron Transport from Two-Dimensional Electron Gas to Quantum Dot
}

\author{
Masakazu Muraguchi, Yukihiro Takada$^1$, Shintaro Nomura$^1$, Tetsuo Endoh, Kenji Shiraishi$^1$}

\inst{%
Center for Interdisciplinary Research, Tohoku University, Sendai, Miyagi 980-8578, Japan\\
$^1$ Graduate School of Pure and Applied Sciences, University of Tsukuba, Tsukuba, Ibaraki 305-8571, Japan
}

\recdate{\today}

\abst{%
 We have theoretically investigated the time-evolution of electron transport from a two-dimensional electron system (2DES) to a quantum dot (QD). We clearly showed that the coherent electron transport is remarkably modified depending on the initial electronic state in the 2DES. The electron transport from the 2DES to the QD is strongly enhanced when the initial state of electron in the 2DES is localized below the QD. We indicate that these features should be obtained in other nano-structure-electrode coupled systems, and the electron transport between the systems having completely different geometrical conditions is significantly affected by the initial spatial distribution of the electron density. 
}

\kword{%
Quantum dot, Two-dimensional electron system, Quantum dynamics, Electron transport
}

\begin{document}
\maketitle

\section{Introduction}

The transport properties of electron from a two-dimensional electron system (2DES) to a quantum dot (QD) have been one of the hottest topics for a couple of decades. Many important results are obtained from a wide variety of scopes \cite{A1,A3,A4,Eto,A5}. Most studies have investigated the electronic state of QD due to the well-defined controllable states of QD. In this paper, we have focused on the electronic state of 2DES. Conventionally, the electronic state in 2DES has not drawn much attention compared to the electronic states of QD in the transport problem. Because, a 2DES has been treated as a heat bath, where most studies assume that the 2DES satisfies an ideal equilibrium stable state.
 However, several studies have reported that 
 the electrons in the metal-oxide-semiconductor field effect transistor (MOSFET) cannot enter contacts without reflection, and the maximum transmission probability is limited \cite{Natori,Datta}. These reports imply that the contacts in MOSFET should not be treated in thermal equilibrium. This argument is also applicable to the 2DES-QD coupled system. The electronic state in the 2DES would influence the electron transport from 2DES to QD.  Therefore, to study the effect of the electronic state in the 2DES is crucial for the precise control of electron transport from 2DES to QD. 

In this paper, we particularly have focused on the geometrical structure of the QD-2DES coupled system, where the 2DES is spread over two dimensionally, whereas the area of QD covers only part of the 2DES. We clarified how the initial spatial distribution of electronic state in the 2DES affects the electron transport from the 2DES to the QD.

\section{Model and Method}
In order to extract the elementary step of electron transport from a 2DES to a QD, we investigated a simple model which a Quantum Dot (QD) weakly couples with a two-dimensional electron system (2DES). 
By applying our approach, that is the numerical finite difference method to solve the time-dependent Schr\"{o}dinger equation (TDSE) to nano-devices\cite{Muraguchi, Muraguchi2, Okunishi}, we obtain the explicit trace of the electron transport process from the 2DES to the QD.

\begin{figure}
\begin{center}
\includegraphics[scale=0.59]{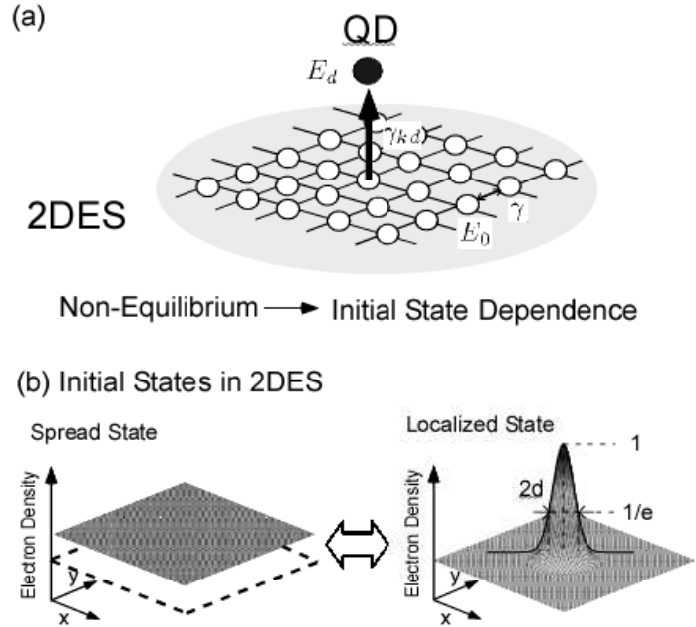}
\caption{
(a) Schematic illustration of the two-dimensional tight-binding lattice (open circle) with a weakly coupled additional site (closed circle), which represent the quantum dot (QD)- two-dimensional electron system (2DES) coupling system.
(b) Schematic illustrations of the initial spatial distribution of electron density in 2DES. The electron density distribution of spatially spread state (left) and spatially localized state (right) are shown. }
\label{f2}
\end{center}
\end{figure}

We employ a tight-binding approximation, where a 2DES and a QD are represented by the two-dimensional periodic tight binding lattice and a weakly coupled additional site, respectively as shown in Fig.\ref{f2}(a).
The detailed conditions of the calculations are the same as our previous study\cite{Muraguchi2}:$\gamma=1.0$, 
$\gamma_{\rm kd}=0.1\gamma$, $E_{\rm k}=0.0$, $E_{\rm d}=-4.0\gamma$ and $N=255\times255$ are chosen.
We have scaled the values of the on-site energies and the transfer integrals by $\gamma$, and time is scaled by $1/\gamma$. 
We assume that the QD is coupled with the tight-binding lattice at time$=0$. In this work, we neglect the electron-electron interaction for simplification, and the time-evolution of an electron is traced.

We choose two types of initial spatial distributions to clarify the initial state dependence of electron dynamics in the present system as shown in Fig.\ref{f2}(b). One is a spatially spread state $[\psi(x,y,0)=\frac{1}{\sqrt{N}}\exp(ik_x x+ik_y y)$, $k_x=k_y=0$], and the other is a spatially localized state (wave packet) [$\psi(x,y,0)=\frac{1}{d\sqrt{\pi}}\exp(-[(x-x_0)^2+(y-y_0)^2]/2d^2)$], where we set the center position of the electron distribution $(x_0,y_0)$ just below the QD. The strength of localization is determined by the width of the wave packet ($d$). Here, we assume that the electron density of an electron wave packet decays $1/e$ from the peak with $d$ ranging from $30$ to $80$ sites.
Then, we calculate the time-evolution of an electron wave function under the above initial conditions.

\section{Results}

\begin{figure}
\begin{center}
\includegraphics[scale=0.6]{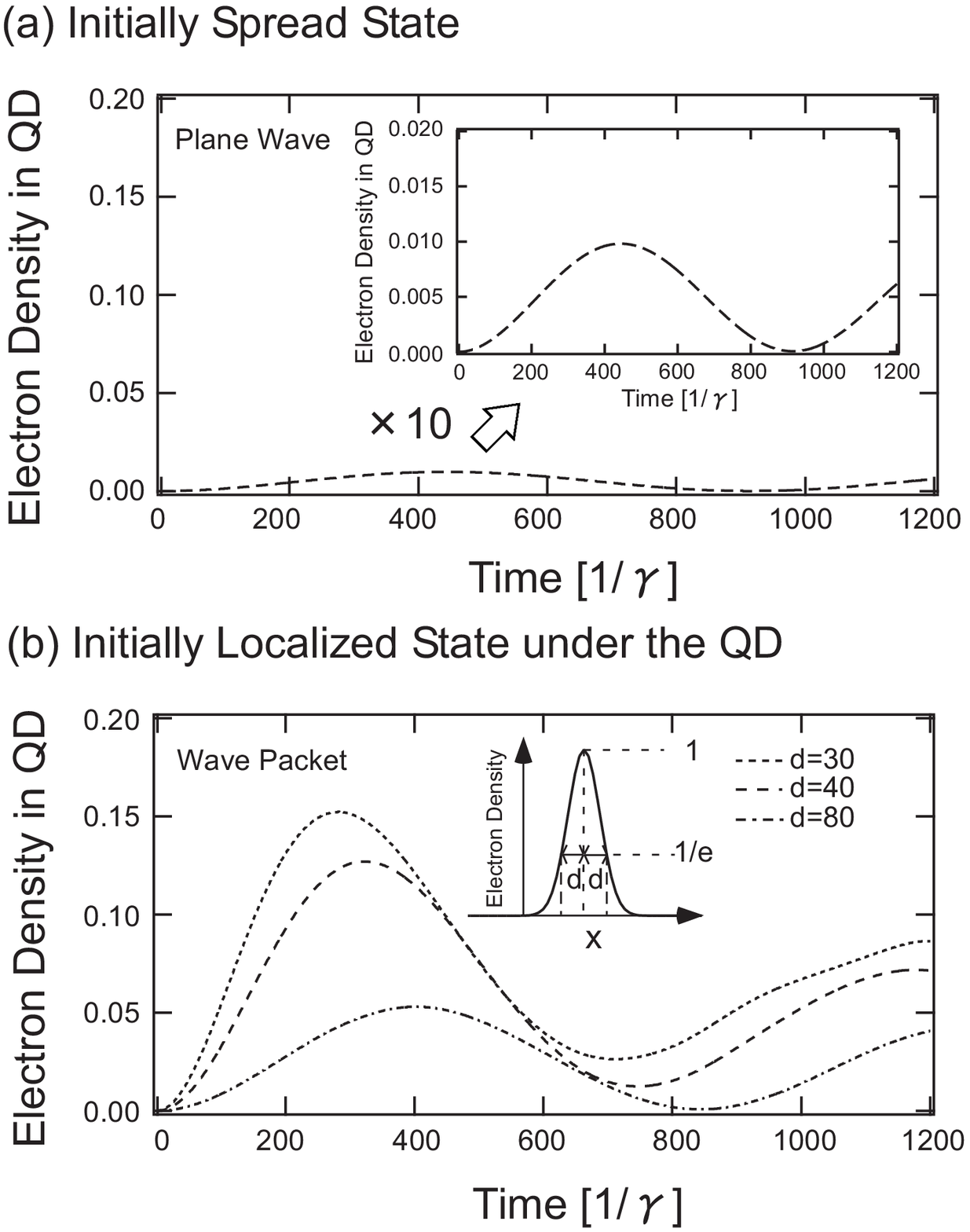}
\end{center}
\caption{The electron density in the QD against time for (a) the initially spread state and (b) the initially localized states. An electron initially in the 2DES transports to the QD in time. The maximum electron density in QD of $d=30$ is about $15$ times larger than the spread state. 
}
\label{f3}
\end{figure}

Figure \ref{f3} indicates the electron density in the QD against time. These results show the dependency of the electron transport to the QD on the above initial conditions. In all conditions, the electron density in the QD oscillates during the time-evolution, which means that the electron transported from 2DES to the QD returns to the 2DES with a certain period of time. However, this oscillation dissipates with passing through time, caused by the spreading of the portion of the electron density over the 2DES. Note that the period of the electron transport between the QD and the 2DES is determined according to the localization strength of the initial electron density. Here, the system is consisted of the QD with a single level and the 2DES with many levels. This system is more complex than that of the simple two-level system, in which the period of the electron transport does not depend on the initial condition. 
The period of the electron transport between them are determined by the energy difference between the QD level and each state in the 2DES. Therefore, it changes with varying the initial electronic state. 

The noticeable point is that the amount of electron density transported from the 2DES to the QD is remarkably different depending on the initial geometrical conditions of the electron spatial distribution. 
 When the initial electron density is spatially spread out, the amount of the electron density transported from the 2DES to the QD is small due to the small overlap between the spread state and the QD. [Fig. \ref{f3}(a)]. Note that the electron density distribution has a dip below the position of the QD and this modulation oscillatory spreads outwards with time. We study this type of modulation in our previous work \cite{Muraguchi2}. 
On the other hand, only when the initial distribution of electron is spatially localized below the position of the QD, the electron transport from the 2DES to the QD is strongly enhanced [Fig. \ref{f3}(b)]. The maximum electron density in the QD at $d=30$ is about $15$ times larger than the initial condition which is spread over the 2DES.
 Figure \ref{f3}(b) also indicates that the electron transport enhances with increasing the localization strength. The maximum electron density in the QD is $3$ times different between the condition $d=30$ and $80$.


\section{Discussion}

\begin{figure}
\begin{center}
\includegraphics[scale=0.65]{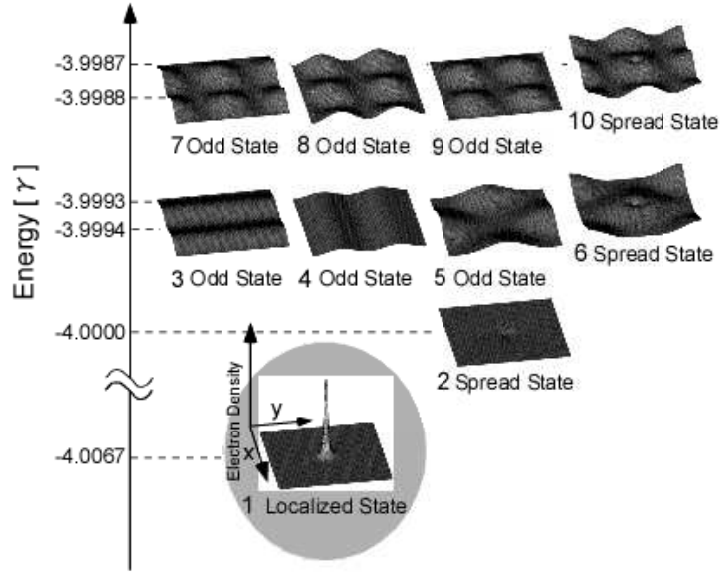}
\end{center}
\caption{The electron density distributions in 2DES of the eigenstates for the condition $\gamma_{\rm kd}=0.1\gamma$ and $E_{\rm d}=-4.0\gamma$. 
The vertical axis shows the energy of the eigenstate.
We classify these eigenstates into three groups according to their spatial distributions of electron density.  
}
\label{f6}
\end{figure}

To obtain a more precise picture of the time-evolution of this system, we employ a projection analysis, where the initial wave function [$\psi(x,y,0)$] is expanded using the eigenstates [$\phi_m(x,y)$] under the condition $\gamma_{\rm kd} =0.1\gamma$. A projection probability of initial states ($P_{m}$) can be described as\begin{eqnarray}
P_{m}=\Bigl|\int 
\phi_{m}^{\ast}(x,y) \psi(x,y,0) dxdy\Bigr|^2.
\end{eqnarray}
We expect that the value of $P_m$ and the spatial density distributions of each eigenstate $\phi_m$ enable us to give detailed features of the time-evolution of the electron in the QD-2DES coupled system.
Figure \ref{f6} shows the electron density distributions of the eigenstates ($\phi_m$) for the QD-2DES coupled system. 
We classified these states into three groups according to their spatial distributions of electron density. We assign the eigenstate spreading over the 2DES as a ``spread state", the one that is localized under the position of the QD as a ``localized state", and the one that does not mix with the QD as the ``odd states". 

We first consider the case which the initial wave function of the electron spread over the 2DES. In this case, the coefficients of the basis functions (eigenstates of the QD-2DES coupled system) are determined by satisfying the uniform distribution of electron density. Accordingly, this initial state is mainly projected to the spatially spread states.  Thus, the time-evolution of the electron density distribution inevitably shows the modulation of electron density caused by the fact that each eigenstate has a different phase velocity in the time-evolution. However, the total amount of electron transported from the 2DES to the QD is limited, because the component of the QD is rather small in these eigenstates.
 
We next consider the case which the initial electronic state in 2DES is spatially localized. We found that this initial state is mainly expanded by the localized state (state 1). 
This state plays a crucial role to induce the electron transport between the 2DES and the QD. Because this localized state has a large density in the QD due to the fact that this state is originally from the localized state of the QD in the condition $\gamma_{\rm kd}=0$. Thus, the electron transport to the QD is strongly enhanced. We also find that the contribution of the localized state becomes larger with increasing the localization. 
These results imply that the tunneling current from 2DEG to QD depends on the temperature of 2DEG, because electron is expected to localize easier in the higher temperature.

We also consider the case which an electron is initially in the QD. The time-evolution of the electron density distribution in the 2DES clearly show that the electron transported from the QD forms a localized distribution in the 2DES. Then, this localized distribution of electron is gradually spread over the 2DES; however a portion is returned to the QD after time passes.
This result is understood by the inverse process of the case with the electron initially localized below the QD in the 2DES.
The reason is simple; this initial state is mainly expanded with the localized state which induces electron transport from the QD to the 2DES.

These results indicate that we should control not only the energy but also the geometrical matching between the peak position of the initial density distribution in the 2DES and the position of the QD.
This feature is derived from our simplified model; however, the characteristics of the initial density distribution dependence can be projected to a generalized system. The electron transport between the systems with completely different geometrical conditions is significantly affected by the initial spatial distribution of the electron density.

\section{Conclusions}

We theoretically studied the time-evolution of electron transport process from 2DES to QD. We clearly showed that the coherent electron transport is modified remarkably depending on the initial electronic state of the 2DES. When the initial state is spatially spread over the 2DES, the electron transport to the QD is small due to the small overlap of the wave function between the QD and the 2DES. On the other hand, when the initial state is localized below the QD, the electron transport to the QD is strongly enhanced with increasing the strength of the localization. 
We pointed out that the similar results should be obtained in other nano-structure-electrode coupled systems and the electron transport between the systems having completely different geometrical conditions is significantly affected by the initial spatial distribution of the electron density. 
We expect that controlling the electronic states in the electrodes would have a potential to give further development to the future of the field of nano-electronic devices.

\acknowledgement
This work was supported by a Grant-in-Aid for Scientific Research No. 18063003 and No. 20760019 from the Ministry of Education, Culture, Sports, Science and Technology, Japan. Computations were carried out of the supercomputer centers of the Institute of Molecular Science and the Institute for Solid State Physics, University of Tokyo.


\begin{thebibliography}{19}
\bibitem{A1}
Y. Takahashi, M. Nagase, H. Namatsu, K. Kurihara, K. Iwadate, Y. Nakajima, S. Horiguchi, K. Murase and M. Tabe: IEEE Electronics Lett. {\bf 31} (1995) 136. 
\bibitem{A3}
A. Bachtold, P. Hadley, T. Nakanishi, C. Dekker: Science {\bf 294} (2001) 1317. \bibitem{A4}
R. M{\"u}ller, J. Genoe and P. Heremans: Appl. Phys. Lett. {\bf 88} (2006) 242105. 
\bibitem{A5}
H. Ishii and T. Nakayama: Phys. Rev. B {\bf 73} (2006) 235311.
\bibitem{Eto} M.Eto, Y.Nazarov, Phys. Rev. B {\bf 66} (2002) 153319 

\bibitem{Natori}
K. Natori, J. Appl. Phys. 76, 4879 (1994).
\bibitem{Datta}
S. Datta, Superlattice and Microstruct. 23, 771 (1998).
\bibitem{Okunishi}
T. Okunishi, M. Muraguchi, and K. Takeda: Phys. Rev. B {\bf 75} (2007) 245314.
\bibitem{Muraguchi2}
M. Muraguchi, Y. Takada, S. Nomura, and K. Shiraishi : Jpn. J. Appl. Phys. {\bf 47} (2008) 7807.
\bibitem{Muraguchi}
M. Muraguchi and K. Takeda: Jpn. J. Appl. Phys. {\bf 46} (2007) 1224.
\end{thebibliography}
\end{document}